\documentclass[12pt]{article}
\newcommand{\nc}{\newcommand}
\nc{\fh}{\hat{f}}
\nc{\muh}{\hat{\mu}}
\nc{\nuh}{\hat{\nu}}
\nc{\bib}{\bibitem}
\nc{\al}{\alpha}
\nc{\g}{\gamma}
\nc{\G}{\Gamma}
\nc{\D}{\Delta}
\nc{\eps}{\epsilon}
\nc{\la}{\lambda}
\nc{\La}{\Lambda}
\nc{\var}{\varphi}
\nc{\cg}{{\cal G}}
\nc{\pa}{\partial}
\nc{\nn}{\nonumber \\ }
\nc{\hf}{\frac{1}{2}}  
\nc{\dz}{\frac{dz}{2\pi i}}
\nc{\bin}[2]{\left (\begin{array}{c} {#1}\\ {#2} \end{array}\right )}
\nc{\ben}{\begin{equation}}
\nc{\een}{\end{equation}}
\nc{\bea}{\begin{eqnarray}}
\nc{\eea}{\end{eqnarray}}
\nc{\bra}[1]{\langle {#1}|}
\nc{\ket}[1]{|{#1}\rangle}
\newcommand{\Z}{\mbox{$Z\hspace{-2mm}Z$}}
\nc{\C}{\mbox{\hspace{1.24mm}\rule{0.2mm}{2.5mm}\hspace{-2.7mm} C}}
\nc{\Nat}{\mbox{\hspace{.04mm}\rule{0.2mm}{2.8mm}\hspace{-1.5mm} N}}

\nc{\HH}{\mbox{\hspace{.04mm}\rule{0.2mm}{2.8mm}\hspace{-1.5mm} H}}

\def\vvdots{\mathinner{\mkern1mu\raise1pt\vbox{\kern7pt\hbox{.}}\mkern2mu
 \raise4pt\hbox{.}\mkern2mu\raise7pt\hbox{.}\mkern1mu}}

\begin{document}

\topmargin -5mm
\oddsidemargin 5mm

\begin{titlepage}
\setcounter{page}{0}

\vspace{8mm}
\begin{center}
{\huge On conformal Jordan cells of finite}\\[.4cm]
{\huge and infinite rank}

\vspace{15mm}
{\Large J{\o}rgen Rasmussen}\\[.3cm] 
{\em Centre de recherches math\'ematiques, Universit\'e de Montr\'eal}\\ 
{\em Case postale 6128, 
succursale centre-ville, Montr\'eal, Qc, Canada H3C 3J7}\\[.3cm]
rasmusse@crm.umontreal.ca

\end{center}

\vspace{10mm}
\centerline{{\bf{Abstract}}}
\vskip.4cm
\noindent
This work concerns in part 
the construction of conformal Jordan cells of infinite rank and 
their reductions to conformal Jordan cells of finite rank.
It is also discussed how a procedure similar to Lie algebra contractions
may reduce a conformal Jordan cell of finite rank to one of lower rank.
A conformal Jordan cell of rank one corresponds to a primary field.
This offers a picture in which any finite conformal Jordan cell of a given
conformal weight may be obtained from a universal covering cell
of the same weight but infinite rank.
\\[.5cm]
{\bf MSC (2000):} 81T40\\[.1cm]
{\bf Keywords:} Logarithmic conformal field theory, Jordan cells.
\end{titlepage}
\newpage
\renewcommand{\thefootnote}{\arabic{footnote}}
\setcounter{footnote}{0}


Logarithmic conformal field theory is essentially based on the appearance of
conformal Jordan cells, or Jordan cells for short, in the spectrum of fields.
We refer to \cite{Gur}
for the first systematic study of logarithmic conformal field theory, and to
\cite{Flo,Gab,Nic} for recent reviews on the subject. 
The number of fields making up a Jordan cell is called the rank
of the cell. Only Jordan cells of {\em finite} rank have been discussed 
in the literature. An objective of this work is to propose an extension 
to Jordan cells of {\em infinite} rank. 

We shall also discuss how Jordan cells of finite or infinite rank may
be reduced to Jordan cells of lower rank.
In this regard, a Jordan cell of rank one merely corresponds to
an ordinary primary field.
The reductions of Jordan cells of finite rank are governed by
a procedure resembling Lie algebra contractions. A related procedure
has recently been employed in \cite{BS,Rabat,Ras,Ras2}, and in its
various guises it constitutes
a promising new approach to logarithmic conformal field theory.
\\[.1cm]
%
%
\mbox{}

A Jordan cell of rank $r=\rho+1$ consists of a primary field, $\Psi_0$, and
$\rho$ logarithmic and non-primary 
partners, $\Psi_1,...,\Psi_\rho$, satisfying \cite{RAK}
\ben
 T(z)\Psi_{j}(w)\ =\ 
 \frac{\D\Psi_{j}(w)+(1-\delta_{j,0})\Psi_{j-1}(w)}{(z-w)^2}
 +\frac{\pa_w\Psi_{j}(w)}{z-w},\ \ \ \ \ \ \ j=0,1,...,\rho
\label{r}
\een
where $\Psi_{-1}\equiv0$. $T$ is the energy-momentum tensor whose
modes, $L_n$, generate the Virasoro algebra
\ben
 [L_n,L_m]\ =\ (n-m) L_{n+m}+\frac{c}{12}n(n^2-1)\delta_{n+m,0}
\label{vir}
\een
with central charge $c$. In terms of these, (\ref{r}) reads
\ben
 [L_n,\Psi_j(z)]\ =\ \{z^{n+1}\pa_z+\D(n+1)z^n\}\Psi_j(z)+(n+1)z^n\Psi_{j-1}(z)
\label{LPsi}
\een
The two-point functions of the fields comprising the Jordan cell are
\bea
 \langle\Psi_i(z)\Psi_j(w)\rangle&=&0,\ \ \ \ \ \ \ \ i+j<\rho\nn
 \langle\Psi_i(z)\Psi_\rho(w)\rangle&=&\frac{\sum_{m=0}^i\frac{(-2)^m}{m!}
   A_{i-m}\left(\ln(z-w)\right)^m}{(z-w)^{2\D}}\nn
 \langle\Psi_i(z)\Psi_j(w)\rangle&=&
  \langle\Psi_{i+j-\rho}(z)\Psi_\rho(w)\rangle\nn
  &=&\frac{\sum_{m=0}^{i+j-\rho}\frac{(-2)^m}{m!}
   A_{i+j-\rho-m}\left(\ln(z-w)\right)^m}{(z-w)^{2\D}},
  \ \ \ \ \ \ \ i+j\geq\rho
\label{twopoint}
\eea
with structure constants $A_j$, $j=0,...,\rho$. As already mentioned, 
it is seen that a rank-one Jordan cell is simply a primary field.

For later convenience, we now 
re-express (\ref{r}) using the following self-explanatory matrix notation
\ben
 T(z)\Psi(w)\ =\ 
  \frac{\D\Psi(w)+P\Psi(w)}{(z-w)^2}
 +\frac{\pa_w\Psi(w)}{z-w}
\label{rP}
\een
where we have introduced the off-diagonal $r\times r$ matrix
\ben
 P\ =\ \left(\begin{array}{ccccccc} 0&0&0&...&0&0&0\\
   1&0&0&...&0&0&0\\   0&1&0&...&0&0&0\\  \vdots&&&\vdots&&&\vdots\\
   \vdots&&&\vdots&&&\vdots\\
    0&0&0&...&1&0&0\\  0&0&0&...&0&1&0  \end{array}\right)
\label{P}
\een
such that $P_{i,j}=\delta_{i,j+1}$. To emphasize the size of a matrix,
we shall sometimes write $P_{r\times r}$ 
or $\Psi_{r\times1}$, for example.

Our first objective is to discuss how a Jordan cell of rank $r=\rho+1$
may be reduced to a Jordan cell of rank $r'<r$.
If the difference $r-r'=2q$ is even, the reduction is straightforward as
one may simply set 
\ben
 \Psi_0\ \equiv\ ... \ \equiv\ \Psi_{q-1}\ \equiv\ \Psi_{\rho-q+1}\ 
  \equiv\ ...\ 
  \equiv\ \Psi_\rho\ \equiv\ 0
\label{equiv0}
\een
The surviving fields, $\Psi_q$, ..., $\Psi_{\rho-q}$, are
seen to form a Jordan cell of rank $r'$ with $\Psi_q$ as its only
primary member. It should be stressed that we in this work confine
ourselves to considering Jordan cells from the point of view
of conformal structure (\ref{r}) and 
(chiral) two-point functions (\ref{twopoint}).
It is immediately more complicated to describe the reduction
if $r-r'$ is odd. We shall focus on the case where this difference
is one as all reductions may be constructed as concatenations
of such unit reductions. This applies as well, of course, to the even case
discussed above, but corresponds to an alternative scenario. 

Let us consider the linear map 
\ben
 \left(\begin{array}{c} \Phi_{0;\eps}\\ \vdots \\ \Phi_{\rho;\eps}
   \end{array}\right)\ =\ 
 \left(\begin{array}{ccc} U_{0,0;\eps}&...&U_{0,\rho;\eps}\\
   \vdots&&\vdots\\ U_{\rho,0;\eps}&...&U_{\rho,\rho;\eps}
   \end{array}\right)\left(\begin{array}{c} \Psi_{0}\\ \vdots \\ \Psi_{\rho}
   \end{array}\right)
\label{U}
\een
of the original representation of the Jordan cell given
in terms of $\Psi$ to a new one in terms of $\Phi_\eps$.
The map is governed by the matrix $U_\eps$ which is assumed 
invertible for non-vanishing $\eps$.
We shall be interested in the limit\footnote{Note that $\Phi_j$ denotes
a single field whereas $\Phi_\eps$ indicates a vector of fields denoted
$\Phi_{j;\eps}$.
This is done to keep the notation simple, and will hopefully not
cause confusion.}
\ben
 \Phi_j(z): =\ \lim_{\eps\rightarrow0}\Phi_{j;\eps}(z)
\label{Phi}
\een
and look for a map (\ref{U}) that would result in a decoupling of
the set $\Phi_0,...,\Phi_\rho$ into a Jordan cell of rank $\rho$
consisting of the first $\rho$ fields, $\Phi_0,...,\Phi_{\rho-1}$, and
a primary field, $\Phi_\rho$, with vanishing two-point function.

To this end, we first consider
\bea
 T(z)\Phi(w)&=&\lim_{\eps\rightarrow0}\{T(z)\Phi_\eps(w)\}\nn
  &=&\frac{\D\Phi(w)+\lim_{\eps\rightarrow0}\{U_\eps PU^{-1}_\eps
    \Phi_\eps(w)\}}{(z-w)^2}+\frac{\pa_w\Phi(w)}{z-w}
\label{TPhi}
\eea
from which it follows that we should require that
\ben
 \lim_{\eps\rightarrow0}\{U_{r\times r;\eps} 
   P_{r\times r}U^{-1}_{r\times r;\eps}\}\ =\ 
  \left(\begin{array}{cc} P_{\rho\times\rho}&0_{\rho\times1}\\ \\
   0_{1\times\rho}&0_{1\times1}\end{array}\right)
\label{UPU}
\een
Likewise, from 
\ben
 \langle\Phi_i(z)\Phi_j(w)\rangle\ =\ \lim_{\eps\rightarrow0}
 \langle\Phi_{i;\eps}(z)\Phi_{j;\eps}(w)\rangle
\label{twophi}
\een
it follows that we should impose the conditions
\bea
 &&\lim_{\eps\rightarrow0}\left\{\sum_{k,l=0}^\rho
  U_{i,k;\eps}U_{j,l;\eps}\sum_{m=0}^{k+l-\rho}
  \frac{(-2)^m}{m!}A_{k+l-\rho-m}(\ln(z-w))^m\right\}\nn\nn
 &&\hspace{1.4cm}=\ \left\{\begin{array}{lc}\sum_{m=0}^{i+j-\rho+1}
  \frac{(-2)^m}{m!}A_{i+j-\rho+1-m}(\ln(z-w))^m,
    \hspace{1cm}&0\leq i,j<\rho\\ \\ 0,&i=\rho\ {\rm or}\ j=\rho
   \end{array}\right.
\label{twoUU}
\eea

We have found the following solution for the matrix $U_{r\times r;\eps}$:
\bea
 U_{i,j;\eps}&=&0,\hspace{1.5cm} i+1<j\nn
 U_{i,i+1;\eps}&=&\frac{\eps}{2}\nn
 U_{i,j;\eps}&=&
  \frac{(-1)^{i-j}}{i-j+1}\bin{2(i-j)}{i-j}\eps^{-2(i-j)-1},
   \hspace{1.5cm} j\leq i<\rho\nn
 U_{\rho,j;\eps}&=&\delta_{j,0}
\label{Usol}
\eea
which for lower rank reads
\ben
 U_{2\times2;\eps}\ =\ \left(\begin{array}{cc} 
  1/\eps&\eps/2\\ \\ 1&0\end{array}\right),
 \hspace{2cm} 
 U_{3\times3;\eps}\ =\ \left(\begin{array}{ccc} 
   1/\eps&\eps/2&0\\ \\
   -1/\eps^3&1/\eps&\eps/2\\ \\
   1&0&0\end{array}\right)
\label{U23}
\een
and
\ben
 U_{4\times4;\eps}\ =\ \left(\begin{array}{cccc} 
  1/\eps&\eps/2&0&0\\ \\
  -1/\eps^3&1/\eps&\eps/2&0\\ \\
  2/\eps^5&-1/\eps^3&1/\eps&\eps/2\\ \\
  1&0&0&0\end{array}\right)
\label{U4}
\een
Before proving that (\ref{Usol}) indeed constitutes
a solution, it is useful to note that
\ben
 U_{(r-1)\times(r-1);\eps}\ =\ R_{(r-1)\times r,r-1}
  U_{r\times r;\eps}C_{r\times(r-1),r}
\label{RUC}
\een
and 
\ben
 R_{(r-2)\times(r-1),r-1}U_{(r-1)\times(r-1);\eps}\ =\
  R_{(r-2)\times(r-1),1}R_{(r-1)\times r,r}U_{r\times r;\eps}
   C_{r\times(r-1),1}
\label{RRUC}
\een
Here we have introduced the matrix 
\ben
 R_{(n-1)\times n,i}\ =\ \left(\begin{array}{ccc} I_{(i-1)\times(i-1)}&
  0_{(i-1)\times1}&0_{(i-1)\times(n-i)}\\ \\
  0_{(n-1)\times(i-1)}&0_{(n-i)\times1}&I_{(n-i)\times(n-i)}\end{array}\right)
\label{Rmat}
\een
annihilating the $i$th row in a matrix $M_{n\times m}$ by multiplication
from the left. Likewise, annihilation of the $j$th column is 
governed by multiplication from the right by the matrix
\ben
 C_{m\times(m-1),j}\ =\ \left(R_{(m-1)\times m,j}\right)^t\ =\ 
  \left(\begin{array}{cc} I_{(j-1)\times(j-1)}&0_{(j-1)\times(m-j)}\\ \\
  0_{1\times(j-1)}&0_{1\times(m-j)}\\ \\
  0_{(m-j)\times(j-1)}&I_{(m-j)\times(m-j)}\end{array}\right)
\label{Cmat}
\een

One does not have to worry about 
working out explicitly the inverse of $U_\eps$ and
subsequently considering the limit in (\ref{UPU}), since
it turns out that
\ben
 U_{r\times r;\eps}P_{r\times r}\ =\ 
   \left(\begin{array}{cc} P_{\rho\times\rho}&0_{\rho\times1}^{(\eps)}\\ \\
   0_{1\times\rho}&0_{1\times1}\end{array}\right)
   U_{r\times r;\eps},\hspace{1.5cm}
  0_{\rho\times1}^{(\eps)}\ =\ 
  \left(\begin{array}{c} \eps/2\\ \\  0_{(\rho-1)\times1}\end{array}\right)
\label{UPUeps}
\een
from which the condition (\ref{UPU}) follows immediately.
The refinement (\ref{UPUeps}) of (\ref{UPU}) follows itself 
from combining (\ref{RUC}) and (\ref{RRUC}).

Now, the conditions (\ref{twoUU}) are easily verified for
$i=\rho$ or $j=\rho$, so we may focus on $0\leq i,j<\rho$.
Using, among other things, that $k+l\geq\rho$ on the left-hand side 
of (\ref{twoUU}), 
we find that (\ref{twoUU}) is equivalent to demanding
\ben
 \lim_{\eps\rightarrow0}\left\{\sum_{n=N-j-1}^{i+1}U_{i,n;\eps}
  U_{j,N-n;\eps}\right\}\ =\ \delta_{N,i+j+1}
\label{UUd}
\een
The non-trivial part of this reads
\ben
 \sum_{n=1}^N\beta_{N,n}\ =\ S_N
\label{S}
\een
where
\ben
 \beta_{N,n}\ =\ \frac{1}{n(N-n+1)}\bin{2N-2n}{N-n}\bin{2n-2}{n-1},
  \hspace{1.2cm} S_N\ =\ \frac{1}{N+1}\bin{2N}{N}
\label{beta}
\een
To prove (\ref{S}) by induction in $N$, we first verify that it is
satisfied for $N=1$. We then try to find $x_2,...,x_N$
such that
\bea
 \beta_{N+1,1}+x_2\beta_{N+1,2}&=&\frac{S_{N+1}}{S_N}\beta_{N,1}\nn
 (1-x_n)\beta_{N+1,n}+x_{n+1}\beta_{N+1,n+1}&=&
   \frac{S_{N+1}}{S_N}\beta_{N,n},\hspace{1.2cm}n=2,...,N-1\nn
 (1-x_N)\beta_{N+1,N}+\beta_{N+1,N+1}&=&\frac{S_{N+1}}{S_N}\beta_{N,N}
\label{ind}
\eea
since this would imply that
\ben
 \sum_{n=1}^{N+1}\beta_{N+1,n}\ =\ \frac{S_{N+1}}{S_N}
  \sum_{n=1}^N\beta_{N,n}\ =\ S_{N+1}
\label{ss}
\een
as required to satisfy the induction step.
The last equality follows from (\ref{S}) which now serves as 
the induction assumption.
The existence of a solution to the set of linear equations (\ref{ind}) 
is non-trivial since there is one more constraint than free parameters.
It is not hard to verify, though, that the set
\ben
 x_n\ =\  \frac{n(n-1)(3N-2n+4)}{N(N+1)(N+2)}  ,\hspace{1.2cm} n=2,...,N
\label{x}
\een
solves (\ref{ind}) and hence the induction step. 

This completes the proof
that the map (\ref{U}) with $U_\eps$ given by (\ref{Usol}) reduces the Jordan
cell $(\Psi_0,...,\Psi_\rho)$ to the Jordan cell $(\Phi_0,...,\Phi_{\rho-1})$,
in the limit $\eps\rightarrow0$.
By construction, the structure constants of the resulting 
lower-rank Jordan cell
are thereby inherited from the original higher-rank Jordan cell.
This should not be regarded as a limitation of the procedure 
since the constants may be modified by a further re-scaling of the new fields.
\\[.1cm]
%
\mbox{}

The conventional representation of a Jordan cell as given above
in (\ref{r}) and (\ref{twopoint}) does not seem to 
allow a naive extension to infinite rank. This may explain why
Jordan cells of infinite rank so far have managed to elude explicit 
construction. To remedy this, we here suggest to first 
re-label the Jordan cell subtracting $\rho/2$ from the index of the fields:
\ben
 r=\rho+1\ {\rm even:}\ \ \ \ \ \ \ \Psi_{-\rho/2},\ \Psi_{-\rho/2+1},...,\ 
   \Psi_{-1/2},\ \Psi_{1/2},...,\ \Psi_{\rho/2}
\label{NS}
\een
and
\ben
 r=\rho+1\ {\rm odd:}\ \ \ \ \ \ \ \Psi_{-\rho/2},\ \Psi_{-\rho/2+1},...,\ 
   \Psi_{-1},\ \Psi_0,\ \Psi_{1},...,\ \Psi_{\rho/2}
\label{R}
\een
In either case, the field $\Psi_{-\rho/2}$ is the only primary one in the
full Jordan cell.
The strong resemblance of (\ref{NS}) and (\ref{R}) 
to $su(2)$ representations of spin $\rho/2$ will be discussed elsewhere.

The limit where the rank approaches infinity is more natural
in the new setting given by (\ref{NS}) and (\ref{R}). 
Namely, let us consider the two
infinite sets of fields $\{\Psi_j\}$ where either $j\in(\Z+1/2)$ or $j\in\Z$,
corresponding to extensions of the two possibilities (\ref{NS}) 
and (\ref{R}), respectively.
In either case, we have
\ben
 T(z)\Psi_j(w)\ =\ \frac{\D\Psi_{j}(w)+\Psi_{j-1}(w)}{(z-w)^2}
  +\frac{\pa_w\Psi_{j}(w)}{z-w}
\label{TP}
\een
whereas the two-point functions shall be defined by
\bea
 \langle\Psi_i(z)\Psi_j(w)\rangle&=&0,\ \ \ \ \ \ \ \ i+j<0\nn
 \langle\Psi_i(z)\Psi_j(w)\rangle&=&\frac{\sum_{m=0}^{i+j}\frac{(-2)^m}{m!}
   A_{i+j-m}\left(\ln(z-w)\right)^m}{(z-w)^{2\D}},\ \ \ \ \ \ \ i+j\geq0
\label{twoinf}
\eea
Indeed, it may be verified straightforwardly that these two-point functions are
quasi-conformally or projectively invariant in the sense that insertions of
$L_{0,\pm1}$ vanish. It is noted that 
\ben
 \langle\Psi_{i+k}(z)\Psi_{j-k}(w)\rangle\ =\
  \langle\Psi_{i}(z)\Psi_{j}(w)\rangle
\label{k}
\een
for all integer $k$, and that, unlike the finite case, there is no primary
field in such a Jordan cell of infinite rank.

{}From (\ref{twoinf}) it seems natural to define
the conjugate field to $\Psi_j$ as $\Psi_{-j}$. This would ensure that 
two-point functions of conjugate fields would be given by the familiar 
pole structure only:
\ben
 \langle\Psi_j(z)\Psi_{-j}(w)\rangle\ =\ \frac{A_0}{(z-w)^{2\D}}
\label{conj}
\een 
A sensible construction would thus presuppose $A_0\neq0$.
With this definition of conjugate fields, we see that
the two types of infinite Jordan cells may be distinguished by whether
or not they contain a self-conjugate field, corresponding to
$j\in\Z$ or $j\in(\Z+1/2)$, respectively. Due to the similarity with
(\ref{R}) and (\ref{NS}), we shall call the two types odd or even,
respectively. It is stressed, though, that for a given {\em finite} rank there is only
one Jordan cell. It may be represented by $(\Psi_0,..., \Psi_{r-1})$
as in (\ref{r}) and (\ref{twopoint}), or by (\ref{NS}) or (\ref{R}) depending on
(the parity of) the rank.

An infinite Jordan cell as defined above reduces to an ordinary Jordan cell
of finite rank, $r=\rho+1$, if one introduces the parameter $\rho$ and defines
\ben
 \Psi_j\ \equiv\ 0,\ \ \ \ \ \ \ \ \ |j|\geq\rho/2+1
\label{red}
\een
This requires that the original Jordan cell of infinite rank
is of the same type as the parity of the rank of the resulting finite cell.
{}From the reduction procedure above, on the other hand,
we know how to reduce a Jordan cell of finite rank to one of any 
lower rank.
This seems to suggest that there are two possible parent Jordan cells
of infinite rank from which any finite Jordan cell can be obtained,
provided, of course, they are of the same conformal weight.
{}From this perspective, one could choose either of the two types
or proposals and consider it the 'universal covering cell' of a
given conformal weight. A reason for
favouring one type over the other does not present itself immediately.
It is not even clear that any of them should be excluded.
The existence or non-existence
of a self-conjugate field, which constitutes
the characterizing difference between the two types,
may play an important role in some contexts.
\\[.1cm]
%
\mbox{}

In summary, we have found that Jordan cells of a given common conformal
weight are naturally organized in an infinite hierarchy labeled by
their rank. A Jordan cell of rank $r$ thus appears at level $r$ in
the hierarchy. The Jordan
cells are nested in the sense that a Jordan cell of rank $r'$
may be obtained
by reduction of a Jordan cell of rank $r$ provided $r'<r$. Two types of 
Jordan cells of infinite rank have been proposed as the universal covering
cell from which any finite Jordan cell may be obtained by reduction.

Finally, our construction of Jordan cells of infinite rank and their reductions
may be extended to $N=1$ superconformal field theory.
This also applies to the reduction procedure of Jordan cells
of finite rank, and will be addressed elsewhere.
\vskip.5cm
\noindent{\em Acknowledgements}
\vskip.1cm
\noindent  The author thanks Michael Flohr for comments.

\end{document}